\documentclass[aps,prl,twocolumn,letterpaper,10pt,footinbib,superscriptaddress,floatfix,amsfonts,amsmath,amssymb]{revtex4-2}

\usepackage{bm,physics,graphicx}
\usepackage[utf8]{inputenc}
\usepackage[english]{babel}
\usepackage[dvipsnames]{xcolor}
\usepackage{hyperref}

\graphicspath{{./Figures/}}
\newcommand{\req}[1]{Eq.~(\hyperref[#1]{\ref*{#1}})}
\newcommand{\reqs}[1]{Eqs.~(\hyperref[#1]{\ref*{#1}})}
\newcommand{\rref}[1]{(\hyperref[#1]{\ref*{#1}})}
\begin{document}
\title{Persistent current noise in narrow Josephson junctions}

\author{Dushko Kuzmanovski}
\affiliation{Nordita\\
KTH Royal Institute of Technology and Stockholm University\\
Hannes Alfv\'{e}ns v\"{a}g 12, SE-106 91 Stockholm, Sweden
}
\author{Rub\'{e}n Seoane Souto}
\affiliation{Division of Solid State Physics and NanoLund, Lund University, S-22100 Lund, Sweden}
\affiliation{Center for Quantum Devices, Niels Bohr Institute, University of Copenhagen, DK-2100 Copenhagen, Denmark}
\author{Alexander V. Balatsky}
\affiliation{Nordita\\
KTH Royal Institute of Technology and Stockholm University\\
Hannes Alfv\'{e}ns v\"{a}g 12, SE-106 91 Stockholm, Sweden
}
\affiliation{Department of Physics, University of Connecticut, Storrs, Connecticut 06269, USA}


\date{\today}

\begin{abstract}
Josephson junctions have broad applications in metrology, quantum information processing, and remote sensing. For these applications, the electronic noise is a limiting factor. In this work we study the thermal noise in narrow Josephson junctions using a tight-binding Hamiltonian. For a junction longer than the superconducting coherence length, several self-consistent gap profiles appear close to a phase difference $\pi$. They correspond to two stable solutions with an approximately constant phase gradient over the thin superconductor connected by a  $2\pi$ phase slip, and a solitonic branch. The  current noise power spectrum has pronounced peaks at the transition frequencies between the different states in each branch. We find that the noise is reduced in the gradient branches in comparison to the zero-length junction limit. In contrast, the solitonic branch exhibits an enhanced noise and a reduced current due to the pinning of the lowest excitation energy to close to zero energy.
\end{abstract}


\maketitle

\label{sec:Introduction}
\emph{Introduction.} Since the theoretical prediction~\cite{Josephson_PhysLett62} and the first experimental demonstrations~\cite{Anderson_PRL63}, the Josephson effect between superconductors at different superconducting phases has attracted a great interest~\cite{Golubov_RMP04}. The Josephson current between two superconductors has been used to redefine the measurement standards and to measure precisely magnetic fields or electromagnetic radiation~\cite{Fagaly_RSI06}. More recently, it has become one of the successful platforms for quantum information processing and remote sensing \cite{Koch_PRA07,Arute_Nat19}.

When the superconducting contact is smaller than the Fermi wavelength, a limited number of channels dominate the electron transport~\cite{Beenakker_SSP91}. In this regime, Andreev bound states (ABSs) appear at the interface between the two superconductors due to the multiple electron-hole reflections~\cite{Andreev_JETP66}. These states, with energies smaller than the gap, dominate the transport properties of the junction \cite{Golubov_RMP04}. ABSs have been proposed to be used for quantum information processing~\cite{Zazunov_PRL03,Janvier_Science15}. The field of mesoscopic superconductivity has experienced a revival after the proposals for engineering Majorana quasiparticles at the ends of one-dimensional superconductors~\cite{Lutchyn_NatRev2018} (for recent reviews,  see \cite{Sauls_Phyl18,Prada_NatRev20}). 

Despite the nondissipative character of the current between two superconductors, it exhibits fluctuations~\cite{Rogovin_AnPhys74,Blanter_Physrep00} that might hinder their use for applications~\cite{vanHarlingen_PRB04}. These fluctuations are dominated by the multiple Andreev reflections at the interface. At finite bias, the noise to current ratio exhibits integer steps as a function of the applied voltage, reflecting the transference of several charges at a time~\cite{Cuevas_PRL99,Cuevas_PRL03}. In the equilibrium situation, the supercurrent noise is due to thermal agitation being peaked at the frequencies matching the energy difference between the occupied and empty states~\cite{Rodero_PRB96,Souto_PRR2020}. The current fluctuations can be measured by embedding the junction into a superconductor loop, inductively coupled to a resonator \cite{Dassonneville_PRL13,Trif_PRB18,Murani_PRL19}.

In long and thin superconductors compared to the superconducting coherence length, the junction exhibits two stable states which carry opposite supercurrent for a given superconducting phase difference~\cite{Troeman_PRB08,Galaktionov_PRB10,Petkovic_NatCom16}. Transitions between these two states, so-called quantum phase slips, can occur due to thermal fluctuations~\cite{Langer_PR67,McCumber_PRB70} or quantum tunneling~\cite{Giordano_PRL88,Zaikin_PRL97,Golubev_PRB01,Matveev_PRL02,Buchler_PRL04}. They lead to an increase of the electrical resistance of the junction~\cite{Zaikin_PRL97,Golubev_PRB01,Lau_PRL01,Zgirski_PRB08,Halperin_IJMP_10} and a suppression of the supercurrent~\cite{Matveev_PRL02}. Quantum phase slips have been measured in thin superconductors embedded into superconducting rings~\cite{Mooij_NatPhys06,Astafiev_Nat12,Belkin_PRX15}. In some cases, the phase slips are suppressed at sufficiently low temperatures, leading to a multivalued supercurrent that can be used to design long-lived quantum memories~\cite{Ligato_arXiv20}.

The existence of multiple values of the supercurrent is due to a renormalization of the order parameter along the thin superconductor. Therefore, the description of the system requires a self-consistent treatment of the order parameter~\cite{Ivlev_1984,Rodero_PRL94,Yeyati_PRB95,Sols_PRB94}. These methods also predict the existence of a metastable solution characterized by a suppression of the order parameter at the center of the nanowire: a soliton. The formation of this soliton leads to a suppression of the supercurrent through the system~\cite{Ivlev_1984,Rodero_PRL94,Yeyati_PRB95,Sols_PRB94}.

In this work, we calculate the supercurrent fluctuations in long and thin superconducting nanowires in the absence of phase slips at constant superconducting phase difference. Complementary works have studied the current noise in the regime where quantum phase slips occur at constant phases~\cite{Semenov_PRB16,Semenov2018}. We study the noise spectrum for the stable and the solitonic solutions, showing that self-consistent effects lead to a suppression of the noise to current ratio in the system.


\label{sec:Model}
\begin{figure}[!ht]
\includegraphics[scale=0.8]{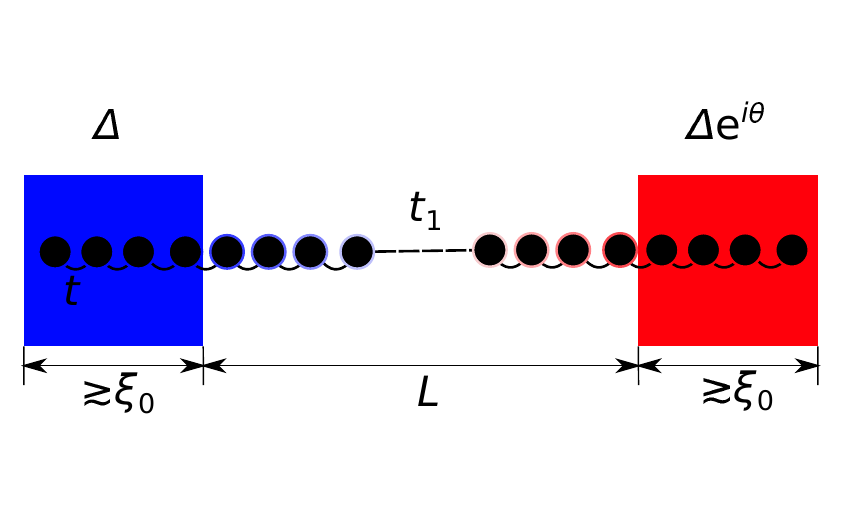}
\caption{\label{fig:sketch}Sketch of the lattice setup. The magnitude $\Delta$ and phase $\theta$ of the SC OP is fixed at the ends of the wire (solid blue and red regions) and relaxed in the winding region of length $L$. The hopping element between the ends of the leads, $t_{1}$, controls the transparency of the junction.}
\end{figure}
\emph{Model.} We model the junction as a flux-biased one-dimensional wire with a discrete number of sites. The wire is superconducting and is described by the mean-field Hamiltonian $\mathcal{H} = \mathcal{H}_{\mathrm{hop}} + \mathcal{H}_{\Delta} - \mu \, \sum_{i} n_{i}$, where:
\begin{subequations}
\label{eq:LeadHams}
\begin{eqnarray}
& \mathcal{H}_{\mathrm{hop}} = -\sum_{i, \sigma} \Bqty{ t_{i} \, e^{-i e \, \Phi_{i, i + 1}} \, c^{\dagger}_{i, \sigma} \, c_{i + 1, \sigma} + \mathrm{H.c.}}, \label{eq:LeadHamKin}
\end{eqnarray}
and $\Phi_{i, i + 1} = \int_{\vb{R}_{i}}^{\vb{R}_{i + 1}} \vb{A} \cdot \dd{\vb{x}}$ is the magnetic flux drop on the bond between sites $i$ and $i + 1$. Assuming that this flux drop is evenly distributed among $L$ bonds in the winding region $\mathcal{W}$, the current that is conjugate to the applied flux is evaluated as a variational derivative over the bond fluxes $I = -\frac{1}{L} \, \sum_{i \in W} \delta \mathcal{H}_{\mathrm{hop}}/\delta \Phi_{i, i + 1}$. We fix the hopping elements to be equal to $t$, except in the middle bond of the winding region where it has a value $t_{1}$, Fig.~\hyperref[fig:sketch]{\ref*{fig:sketch}}, controlling the transparency of the junction, $\tau = \qty[\frac{2 \, t_{1}/t}{1 + \qty(t_{1}/t)^{2}}]^{2}$~\cite{Rodero_PRL94}, and the strength of the inverse proximity effect.

The superconducting Hamiltonian corresponds to an on-site, spin-singlet pairing:
\begin{equation}
\mathcal{H}_{\Delta} = \sum_{i} \Bqty{\Delta_{i} \, c^{\dagger}_{i, \uparrow} \, c^{\dagger}_{i, \downarrow} + \mathrm{H.c.}} + \sum_{i \in \mathcal{W}} \frac{\abs{\Delta_{i}}^2}{U}. \label{eq:LeadHamDelta}
\end{equation}
\end{subequations}
Here, $\Delta_{i}$ are the complex superconducting order parameters (SC OPs) at site $i$ and $U$ is an on-site pairing strength. We note that there is gauge freedom due to the choice of the electron annihilation operators $c_{i\sigma} \rightarrow e^{i e \, \chi_{i}} \, c_{i\sigma}$. It involves a simultaneous transformation of the bond fluxes $\Phi_{i, i + 1} \rightarrow \Phi_{i, i + 1} + \qty(\chi_{i + 1} - \chi_{i})$ and the phases of the SC OPs $\Delta_{i} \rightarrow e^{2 i e \, \chi_{i}} \, \Delta_{i}$. We explore this freedom to fix the flux bonds to zero [multiples of $\pi$; i.e., the magnetic flux quantum $\Phi_{0} = h/(2e)$] and vary the SC OPs in the winding region $\mathcal{W}$. We impose the necessary condition for extremum of the free energy:
\begin{eqnarray}
& 0 = \frac{\delta F}{\delta \Delta^{\ast}_{i}} = \expval{\frac{\mathcal{H}_{\Delta}}{\delta \Delta^{\ast}_{i}}} = \frac{\Delta_{i}}{U} + \expval{c_{i, \downarrow} \, c_{i, \uparrow}}, \ i \in \mathcal{W}. \label{eq:Self-Cons}
\end{eqnarray}
 
After obtaining the site-dependent SC OP $\Delta_{i}$, we can diagonalize the Bogoliubov-de Gennes matrix $\check{H}_{\mathrm{BdG}}$ that enters in the commutator:
\begin{equation}
-\qty[\mathcal{H}, \Psi] = \check{H}_{\mathrm{BdG}} \cdot \Psi, \label{eq:BdGHam}
\end{equation}
with the Nambu spinor $\Psi = \qty(c_{i\uparrow}, c^{\dagger}_{i\downarrow})^{\top}$ describing a spin-singlet pairing. The BdG eigenvectors $\ket{Y^{(n)}} \equiv \qty(u^{(n)}_{i}, v^{(n)}_{i})^{\top}$ correspond to the eigenvalues $E_{n}$, and particle-hole symmetry implies that $\ket{Y^{(-n)}} = \qty(-\qty(v^{(n)}_{i})^{\ast}, \qty(u^{(n)}_{i})^{\ast})^{\top}$ is an eigenvector corresponding to the eigenvalue $-E_{n}$. Then, the expectation value in \req{eq:Self-Cons} can be evaluated as:
\begin{equation}
\expval{c_{i\downarrow} c_{i\uparrow}} = \sum_{n} u^{(n)}_{i} \qty(v^{(n)}_{i})^{\ast} \, n_{\mathrm{F}}\qty(E_{n}), \label{eq:PAexpr}
\end{equation}
where $n_{\mathrm{F}}$ is the Fermi distribution function.

The second-quantized expression of the current operator can be written as a quadratic form $I = \Psi^{\dagger} \cdot \check{J} \cdot \Psi$, with the current operator matrix:
\begin{subequations}
\label{eq:CurMatrix}
\begin{eqnarray}
& \check{J}_{k, l} = \begin{pmatrix}
\mqty{\hat{J}_{k, l}} & 0 \\
0 & -\mqty{\hat{J}_{l, k}}
\end{pmatrix}, \label{eq:CurMatBdG} \\
& \hat{J}_{k, l} = \frac{e}{L} \sum_{i \in \mathcal{W}} \left\lbrace \frac{1}{i} t_{i} e^{-i e \, \Phi_{i, i+1}} \delta_{k, i} \delta_{l, i + 1} \right. \nonumber \\
& \left. - \frac{1}{i} t^{\ast}_{i} e^{i e \, \Phi_{i, i +1}} \delta_{k, i + 1} \delta_{l, i}\right\rbrace, \label{eq:CurMatEl2}
\end{eqnarray}
\end{subequations}
where $k$ and $l$ are site indices.
Using the BdG eigenvectors and eigenvalues, the expectation value of the current operator $\expval{I}$ is:
\begin{eqnarray}
& \expval{I} = 2 \, \sum_{n} \mel{u^{(n)}}{\hat{J}}{u^{(n)}} \, n_{\mathrm{F}}\qty(E_{n}). \label{eq:CurentExpValue}
\end{eqnarray}

At the same time, the current noise spectrum $S\qty(\Omega)$ at finite frequency $\Omega$ is given by the double sums over the eigensystem:
\begin{subequations}
\label{eq:CurrentNoise}
\begin{eqnarray}
& S\qty(\Omega) = \sum_{m, n} \abs{\mel{Y^{(n)}}{\check{J}}{Y^{(m)}}}^{2} \,  W_{m, n}\qty(\Omega), \label{eq:S} \\
& W_{m, n}\qty(\Omega) = \qty[1 - \tanh\qty(\frac{E_{m}}{2 T}) \, \tanh\qty(\frac{E_{n}}{2 T})] \nonumber \\
& \times \frac{\eta}{\qty(\Omega - E_{m} + E_{n})^{2} + \eta^{2}}, \label{eq:Wfunc}
\end{eqnarray}
\end{subequations}
where $\eta$ is a smearing factor for the numerical implementation of a Dirac delta function via a Lorentzian, taken to be much smaller than any other energy scale. The derivation of \req{eq:CurrentNoise} is outlined in the Appendix of Ref.~\cite{Souto_PRR2020}, and the result coincides with the one from Ref.~\cite{Trif_PRB18}.

\label{sec:SelfCons}
\emph{Self-consistent solutions.}
We obtain the SC OP by self-consistently solving \reqs{eq:Self-Cons}, \rref{eq:BdGHam}, and \rref{eq:PAexpr} for each site of the winding region $\mathcal{W}$. In order to make contact with experiment, we express lengths in units of the zero-temperature SC coherence length $\xi_{0}$, and energies in units of the zero-temperature magnitude of the SC OP $\Delta_{0}$. We choose the chemical potential $\mu = 0$, corresponding to half-filling, although our results do not depend on this choice. The pairing strength is adjusted to $U/t = 1.37$ so that $\Delta_{0}/t = 0.0836$, which corresponds to  $\xi_{0}/a  = \frac{\sqrt{\qty(2 t)^{2} - \mu^{2}}}{\Delta_{0}}=24.0$. In order to avoid edge effects due to open boundary conditions, we fix the magnitude and phase of the SC OP to the bulk values on leads with length of the order of $\xi_{0}$ on each end; see Fig.~\hyperref[fig:sketch]{\ref*{fig:sketch}}.

The gradient (stable) branch is obtained by scanning the phase bias, starting from zero bias, with the initial guess for $\Delta_i$ given by the solution of the previous step.
\begin{figure}[!t]
\includegraphics[scale=1.0]{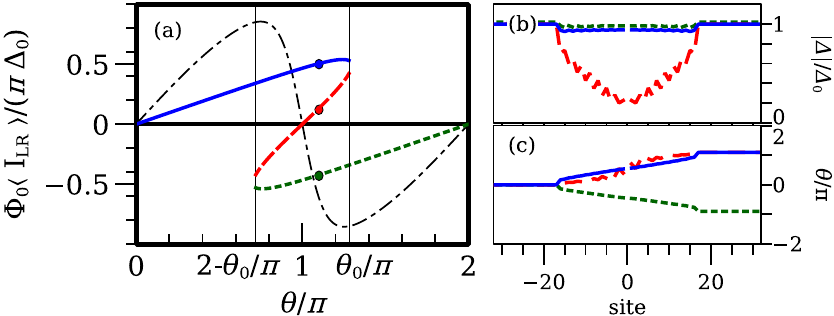}
\caption{(a) Supercurrent as a function of the phase bias for the gradient (solid blue), the solitonic (dashed red) and the gradient after a phase-slip event (dotted green). 
We also show the $L\to0$ result, corresponding to a quantum point contact (dash-dotted gray). (b) SC OP magnitude and (c) phase profiles at the $\theta$ value denoted by the thick point in panel (a). The transparency is $\tau = 1$, the length is $L/\xi = 4/3$, and the temperature is $T/\Delta_{0} = 0.10$. For these parameters, the phase slip occurs at $\theta_{0}/\pi = 1.28$.}
\label{fig:cur}
\end{figure}
We find that the BdG equations have two stable solutions for $L$ greater than but similar to $\xi$ for a superconducting phase difference $2\pi - \theta_0<\theta<\theta_0$ and having a single solution otherwise. At $\theta_{0}$ the system undergoes a discontinuous phase slip, characterized by a change of $2\pi$ on the accumulated superconducting phase in the winding region~\cite{Petkovic_NatCom16}. The value of $\theta_{0}$ becomes closer to $\pi$ for lower values of transparency $\tau$ or lower values of the ratio $L/\xi$. The last case may be achieved either by shortening the leads or by increasing the temperature, which decreases the bulk value $\Delta$ and increases $\xi$. For a phase bias where more than one stable solution to \req{eq:Self-Cons} is possible, a solitonic solution appears. This solution is metastable and the gradient descent method is not able to find it. For this reason, we use Broyden's good Jacobian approximation~\cite{2020SciPy-NMeth, IterMethBook.ch7}. To obtain this solution, we start at phase bias $\pi$ with a real SC OP crossing zero in the middle of the junction.

\label{sec:Results}
\emph{Results.} The BdG equations \req{eq:BdGHam} have three different solutions that coexist close to $\theta=\pi$ for $L$ larger than but similar to $\xi$. This leads to a multivalued tunneling current~\req{eq:CurentExpValue} and SC OP, as shown in Fig.~\hyperref[fig:cur]{\ref*{fig:cur}}.

The current in the two gradient (stable) branches is the same after a phase-slip event occurs, i.e., a shift of $2\pi$ in the superconducting phase difference~\cite{Golubov_RMP04}. This is better illustrated by the order parameter, Figs.~\hyperref[{fig:cur}]{\ref*{fig:cur}(b)} and \hyperref[{fig:cur}]{\ref*{fig:cur}(c)}, where the magnitude of the SC OP is constant in the gradient branches, while the phase acquires an extra $2\pi$ factor. We show that the self-consistent effects on the SC OP lead to a reduction of the current in the gradient branches.
\begin{figure}[!t]
\includegraphics[scale=0.7]{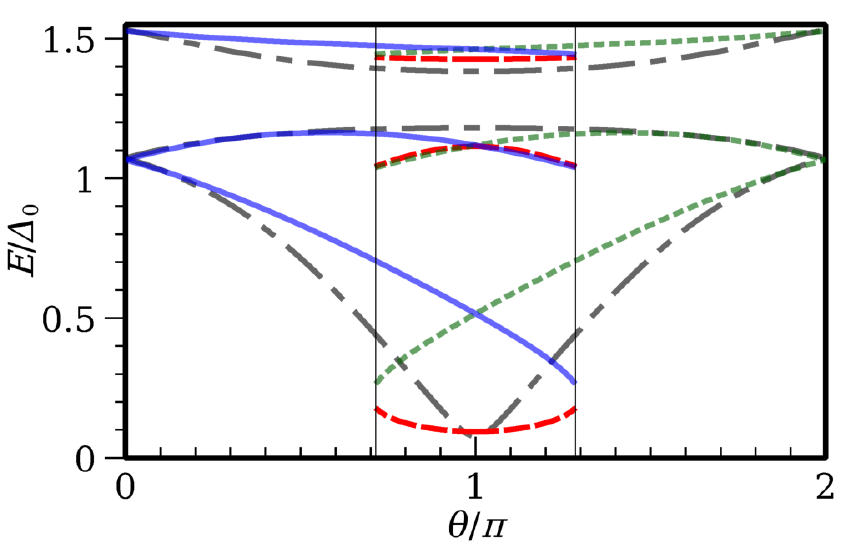}
\caption{Lowest positive eigenvalues $E_{1}$, $E_{2}$, and $E_{3}$, from bottom to top, of the BdG Hamiltonian as a function of phase bias. We show results for the upper gradient (solid blue), solitonic (dashed red), and the lower gradient branches (dashed green). We also show the $L\to0$ limit (dash-dotted gray). The parameters are the same as in Fig.~\hyperref[fig:cur]{\ref*{fig:cur}}.}
\label{fig:abs}
\end{figure}
Also, the current in these solutions is maximal at $\theta_0\geq\pi$, being finite at $\pi$. This is in contrast to the current in short atomic contacts~\cite{Golubov_RMP04}. We note that a non-self-consistent approach, where the SC OP is taken constant and the phase jumps at the weak link, is not a good solution to the problem for $L>\xi$, as it violates current conservation. Finally, the solitonic branch shows a lower supercurrent, consistent with the suppression of the SC OP towards the center of the wire~\cite{Ivlev_1984,Rodero_PRL94,Yeyati_PRB95,Sols_PRB94}.

The effect of a self-consistent treatment of the problem in the BdG excitation spectrum of the thin superconductor is shown in Fig.~ \hyperref[{fig:abs}]{\ref*{fig:abs}}. For short and high-transmitting junctions, the lowest excitation energy approaches $E=0$ at phase $\pi$. In contrast, the lowest excitation energy in the gradient branches does not approach $E=0$ at $\pi$. It shows instead a higher value, becoming minimal at $\theta_0$, consistent with the behavior of the current. In contrast, the lowest excitation in the solitonic branch is pinned close to zero energy in the whole phase coexistence region. The excitation spectrum indicates that the gradient solution is energetically favorable compared to the non-self-consistent or solitonic solution. Finally, we note the small differences found between the non-self-consistent solution and the other branches for the states with $E>\Delta_0$, indicating that it is a good approximation for the states above the bulk gap.

\begin{figure*}[!ht]
\includegraphics[scale=0.8]{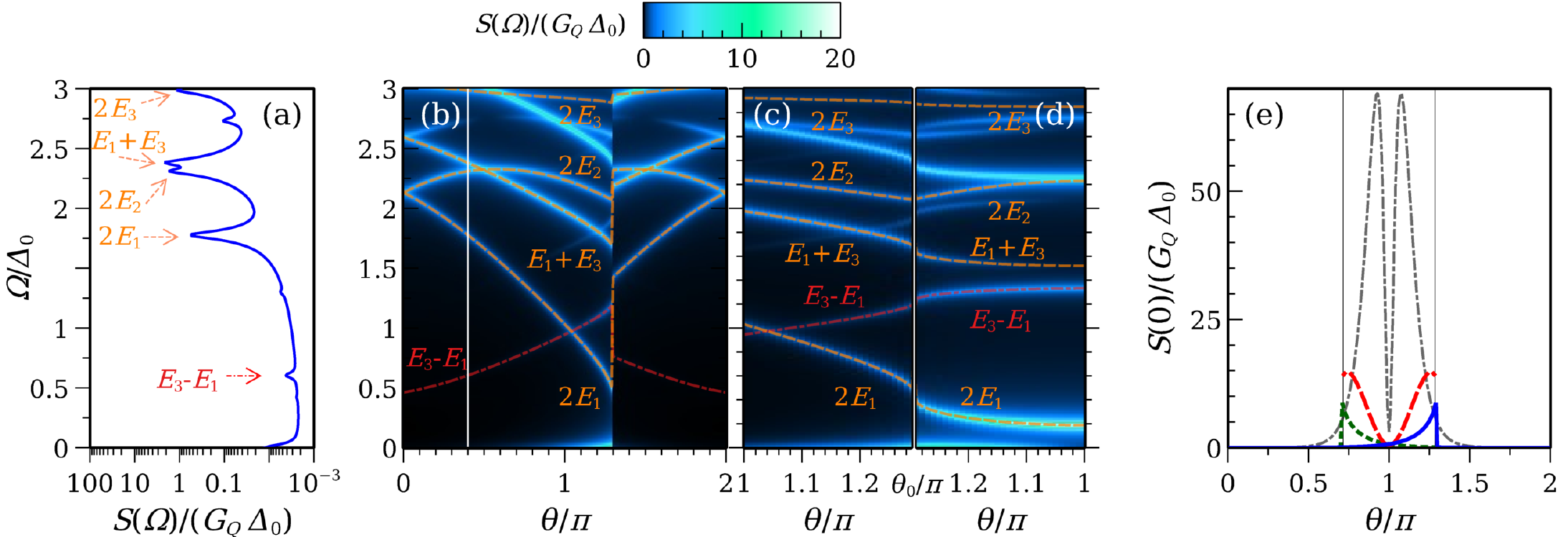}
\caption{\label{fig:noise}(a) Current noise spectrum at a phase bias $\theta/\pi = 0.4$. (b) Density plot of the current noise spectrum as a function of the phase and frequency for the gradient branches. The white vertical line is the phase bias corresponding to panel (a). Panels (c) and (d) show the noise spectrum in the coexistence region for the upper gradient and solitonic branch, respectively. In (b)--(d) we mark the dominant peaks between the eigenvalues with opposite sign (dashed orange) and subdominant peaks between the eigenvalues with equal sign (dotted red) shown in Fig.~\hyperref[fig:abs]{\ref*{fig:abs}}. (e) Zero-frequency noise as a function of the phase for the upper (solid blue) and lower (dotted green) gradient branches, solitonic branch (dashed red), and the $L\to0$ limit (dash-dotted gray). The choice of parameters is the same as in Fig.~\hyperref[fig:cur]{\ref*{fig:cur}}.}
\end{figure*}

The current noise spectrum, \reqs{eq:CurrentNoise}, for the gradient branches is shown in Figs.~\hyperref[fig:noise]{\ref*{fig:noise}(a)} and \hyperref[fig:noise]{\ref*{fig:noise}(b)}. It exhibits a discontinuous jump due to a phase slip at $\theta_{0}$. We note that the noise spectrum for $\theta>\theta_0$ is similar to the one shown for $\theta<2\pi-\theta_0$. The dominant peaks correspond to excitation of two quasiparticle states due to transitions between occupied and empty states, indicated by orange lines in Figs.~\hyperref[fig:noise]{\ref*{fig:noise}(b)}--\hyperref[fig:noise]{\ref*{fig:noise}(d)}. They appear at the sum between two excitation energies in Fig.~\hyperref[fig:abs]{\ref*{fig:abs}} due to electron-hole symmetry~\cite{Kos_PRB13}. There is another set of peaks appearing at the difference between two excitation energies. They correspond to transitions between two excited states, indicated by red lines in Figs.~\hyperref[fig:noise]{\ref*{fig:noise}(b)}--\hyperref[fig:noise]{\ref*{fig:noise}(d)}. The strength of these peaks is proportional to the population of the lower excited state and, therefore, suppressed at zero temperatures. We note that transitions, such as the ones between the first and third to the second BdG eigenstate, are forbidden by selection rules for the current operator, imposed by the symmetry of the model under Andreev reflection relative to the middle point of the junction.

In panels (c) and (d) of Fig.~\hyperref[fig:noise]{\ref*{fig:noise}} we compare the noise spectrum in the gradient and in the solitonic branches. The pinning of the lowest BdG excitation at low energies leads to an increase of the line intensities associated with $E_1$. In particular, we note the increase on the intensity of the line at $E_3-E_1$ due to a finite equilibrium population of the lowest excited state in the BdG spectrum in Fig.~\hyperref[fig:abs]{\ref*{fig:abs}}.

The most transparent comparison of noise in the different branches is in terms of the zero-frequency noise; see Fig.~\hyperref[fig:noise]{\ref*{fig:noise}(e)}. The gradient branches exhibit a lower noise than the $L\to0$ solution, missing the double peak feature around phase $\pi$. The noise becomes appreciable at the coexistence region for the gradient branches, becoming maximal at $\theta_0$ when the lowest BdG excitation approaches zero energy. In contrast, the solitonic branch exhibits an enlarged current noise with respect to the gradient solutions at high transmissions. This is due to the pinning of the lowest BdG excitation to close to zero energy. This increase on the noise to current ratio is another indication of the metastability of the solitonic branch, also shown by the BdG excitation spectrum in Fig.~\hyperref[fig:abs]{\ref*{fig:abs}}.

In the main text, we have focused on the perfect transmission situation. However, the conclusions remain valid for different $\tau$ values, as shown in the Supplemental Material~\footnote{See Supplemental Material [attached at the end] for three figures complementary to Figs. \hyperref[fig:cur]{\ref*{fig:cur}}--\hyperref[fig:noise]{\ref*{fig:noise}} from the
main text, corresponding to a different choice of junction transparency.}. By reducing the transparency, $\theta_0$ tends to approach $\pi$ and the gradient branches become less dependent on the superconducting phase difference. In the limit $\tau\ll1$, self-consistent effects are not important and the solution converges to the non-self-consistent result.

\label{sec:Conclusions}
\emph{Conclusions.} In this work, we have analyzed the current and the current noise through a long and thin junction between two phase-biased superconductors. A self-consistent treatment of the Bogoliubov-de Gennes (BdG) Hamiltonian leads to three different solutions for superconductors longer than their coherence length. The two gradient (stable) solutions show an almost constant superconducting order parameter, while the phase winds smoothly over the whole length of the wire, leading to a finite current at phase difference $\pi$. Above $\pi$, the system exhibits a discontinuous phase slip, a discontinuous jump of $2\pi$ in the phase difference, characterized by a transition between the two gradient branches. The BdG Hamiltonian predicts also the existence of a solitonic (metastable) solution, characterized by a suppression of the superconducting order parameter, reducing the current. This metastable solution shows a pinning of the lowest Bogoliubov excitation energy close to zero energy. We find that self-consistent effects tend to reduce the noise-to-current ratio in the gradient branches.
The noise reduction is due to the increased value of the lowest Bogoliubov excitation energy given by the self-consistent solution with respect to the non-self-consistent one.
In contrast, the low-energy state in the solitonic branch leads to an enhancement of the current noise at finite temperature, due to the excitation of quasiparticles to this state.

We believe that the noise reduction in thin superconductors as a function of their length can be tested in superconducting nanowires, where deterministic phase slips have been observed \cite{Ligato_arXiv20}, using noise spectroscopy techniques \cite{Dassonneville_PRL13,Murani_PRL19}. The reduced noise to current ratio found in these junctions makes thin superconductors an attractive alternative for designing Josephson electronic devices with controlled sources of dissipation and decoherence.

\begin{acknowledgments}
\emph{Acknowledgments.} The authors wish to thank A. Levy-Yeyati for very useful suggestions. This work was supported by VR 03997, the European Research Council under the European Union's Seventh Framework Program Synergy HERO, by the University of Connecticut and the Knut and Alice Wallenberg Foundation KAW 2019.0068. RSS acknowledges the funding from the European Research Council (ERC) under the European Union’s Horizon 2020 research and innovation programme under Grant agreement No.~856526. RSS acknowledges funding from QuantERA project ``2D hybrid materials as a platform for topological quantum computing" and from NanoLund.
\end{acknowledgments}

\bibliography{multibranch_noise}
%
\clearpage
\newpage
\begin{widetext}
\begin{center}
	\textbf{\large Supplemental Material: Persistent current noise in narrow Josephson junctions}
\end{center}
\end{widetext}
\setcounter{equation}{0}
\setcounter{figure}{0}
\setcounter{table}{0}
\setcounter{page}{1}
\makeatletter
\renewcommand{\theequation}{S\arabic{equation}}
\renewcommand{\thefigure}{S\arabic{figure}}
\renewcommand{\bibnumfmt}[1]{[S#1]}

\begin{figure}[!ht]
    \centering
    \includegraphics[scale=1.0]{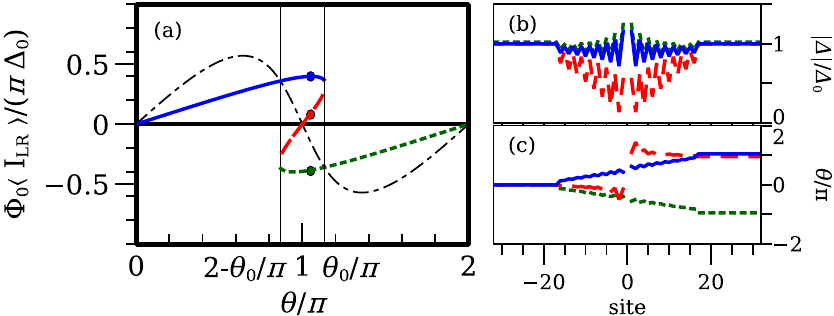}
    \caption{Current (a) and magnitude (b) and phase (c) of the SC OP. We show results for the gradient branches (solid blue and dotted green), the solitonic branch (dashed red) and the non-self-consistent case (black dotted-dashed line). The parameters are the same as in Fig.~\hyperref[fig:cur]{\ref*{fig:cur}} from the main text, with $t_{1}/t = 0.7$.}
    \label{fig:curSM}
\end{figure}

\begin{figure}[!ht]
    \centering
    \includegraphics[scale=0.8]{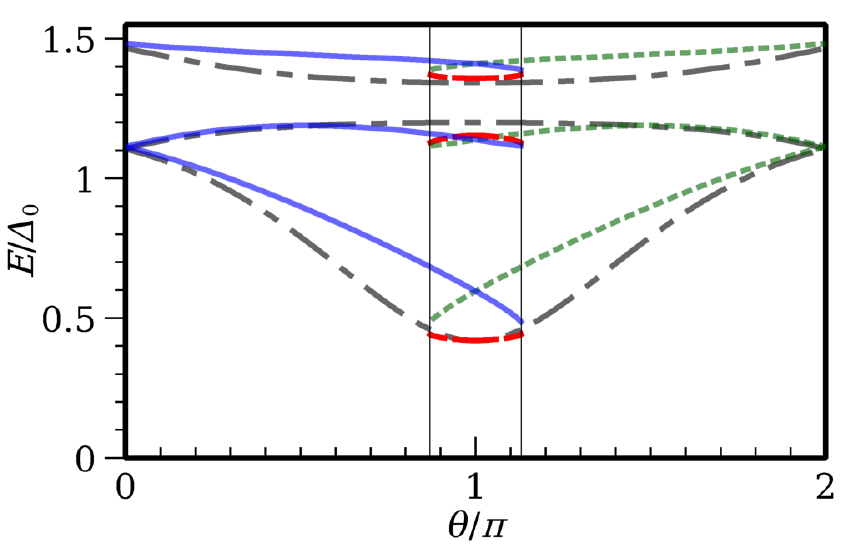}
    \caption{Lowest excitation energies for the same three branches  and parameters as in Fig.~\hyperref[fig:curSM]{\ref*{fig:curSM}}.}
    \label{fig:absSM}
\end{figure}

\begin{figure*}[!ht]
    \centering
    \includegraphics[scale=0.8]{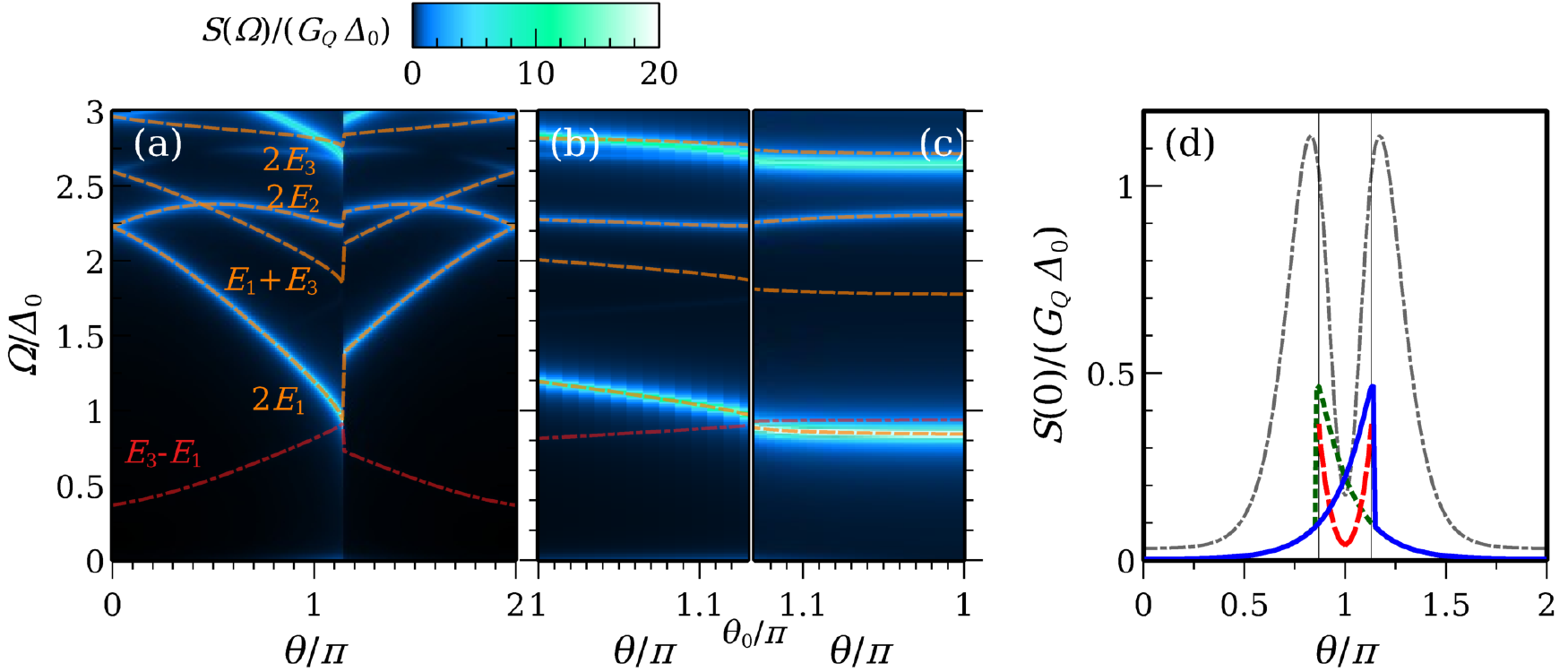}
    \caption{(a) Noise spectrum for the gradient branches. In (b) and (c) we show the noise spectrum in the gradient (b) and solitonic (c) branches in the coexistence region. Panel (d) shows the zero-frequency noise for the same solutions as in Fig.~\hyperref[fig:curSM]{\ref*{fig:curSM}}. Parameters are the same as in Fig.~\hyperref[fig:curSM]{\ref*{fig:curSM}}.}
    \label{fig:noiseSM}
\end{figure*}

\label{sec:SM}
For comparison, we present analogous results in Figs.~\ref{fig:curSM}--\ref{fig:noiseSM} for a junction with a tunnel element $t_{1}/t = 0.7$, corresponding to a transparency of $\tau = 0.88$. In this case, the lowest excitation energy goes to $\sqrt{1 - \tau} = 0.34$ at $\pi$ phase bias for the non-self-consistent solution ($L\to0$ limit). The coexistence region has reduced its size with respect to the highly transmitting situation. In this case we find $\theta_{0}/\pi = 1.13$, indicating the strong sensitivity of the coexistence region to the JJ transparency.

In Fig.~\hyperref[fig:curSM]{\ref*{fig:curSM}} we show the current and the superconducting order parameter (SC OP) for the three different solutions of the BdG equations and the non-self-consistent solution. The reduced transmission with respect to the transparent case (Fig.~\hyperref[fig:cur]{\ref*{fig:cur}} of the main text) leads to a reduced value for the current. We note also an enhancement of the wringing of the magnitude of the SC OP with a non-monotonous winding of the phase in the solitonic branch, Fig.~\hyperref[fig:curSM]{\ref*{fig:curSM}(c)}.

This behavior is accompanied by an increase of the lowest excitation energy, Fig.~\hyperref[fig:absSM]{\ref*{fig:absSM}}. This becomes more evident in the case of the solitonic solution, whose energy is pinned at a higher value in comparison to Fig.~\hyperref[fig:abs]{\ref*{fig:abs}} in the main text. We note that this energy coincides with the lowest excitation value of the non-self-consistent solution in the whole coexistence region, Fig.~\hyperref[fig:absSM]{\ref*{fig:absSM}}.

In Fig.~\hyperref[fig:noiseSM]{\ref*{fig:noiseSM}} we show the current noise for the different branches. We note a decrease on the current noise due to the lower transmission of the junction and the higher energy of the lowest excited states. In particular, we note the disappearance of the noise peak at a frequency $\Omega/\Delta=E_3-E_1$ in the solitonic and the gradient branches due to a higher value of $E_1$.

At zero frequency, we note a lower value of the current noise in the solitonic branch with respect to the gradient ones. This is mainly due to a stronger suppression of the noise in the solitonic branch. In any case, the noise is always smaller than the one predicted for the non-self-consistent solution ($L\to0$ limit). In all the cases the overall zero-frequency noise  level is significantly reduced with respect to the $\tau=1$ case.
\end{document}